\documentstyle[aps,preprint,prb]{revtex}

\begin{document}
\author{R. E. Cohen}
\address{Carnegie Institution of Washington, 5251 Broad Branch RD., N.W., Wahsington\\
D.C. 20015}
\title{Surface Effects in Ferroelectrics: Periodic Slab Computations for BaTiO$_3$}
\date{\today}
\maketitle

\begin{abstract}
Total energies, electronic structure, surface energies, polarization,
potentials and charge densities were studied for slabs of BaTiO$_3$ using
the Linearized Augmented Plane Wave (LAPW) method. The depolarization field
inhibits ferroelectricity in the slabs, and the macroscopic field set up
across a ferroelectric slab is sufficient to cause electronic states to span
the gap and give a metallic band structure, but the band shifts are not
rigid and O $p$ states tend to pile up at the Fermi level. There are
electronic surface states, especially evident on TiO$_2$ surfaces. The
dangling bonds bond back to the surface Ti's and make the surface stable and
reactive. The BaO surfaces are more ionic than the bulk.
\end{abstract}


\section{Introduction}

Surfaces of ferroelectrics can strongly effect their properties. In general
this is a much stronger effect than in paraelectric materials, because the
termination of a ferroelectric gives rise to a depolarization field, which
is huge in an ideal, ionic ferroelectric. The depolarization field, $E_{d%
\text{,}}$ is opposite in sign to the bulk polarization, and gives a
contribution to the energy of $-E_d\cdot P$ large enough to completely
destabilize the bulk ferroelectric state \cite{1,2}. The ferroelectric
distortion in thin films varies with thickness so that the structures become
cubic at thin edges \cite{3}. Huge particle size effects are observed \cite
{4,5}, orders of magnitude greater than observed for non-ferroelectrics.
Surfaces are important also during crystal growth, and their properties are
key to catalysis and use as a epitaxial substrate. Here we study thin slabs
of BaTiO$_3$ to better understand the electronic structure of the surface,
surface energetics and relaxation, and intrinsic screening of the
depolarization field.

Experimental studies of BaTiO$_3$ surfaces are few and many more studies
have focused on SrTiO$_3$ which lacks the complications of ferroelectricity.
Photoemission studies of sputtered and annealed BaTiO$_3$ show evidence of
significant surface defects \cite{6}, and a number of papers have addressed
the defect properties of the surface \cite{7,8,9}. Studies of SrTiO$_{3\text{
}}$surfaces show evidence for minor relaxations by motions of atoms
perpendicular to the surfaces \cite{10,11} and show no evidence for states
in the gap. \cite{12} Some recent studies of BaTiO$_3$ show evidence for gap
states \cite{8}, and others do not \cite{13}, probably indicating
sensitivity to the surface defect state. Photoemission spectroscopy on BaTiO$%
_3$ shows excellent agreement with bulk LAPW computations of the LDA band
structure \cite{13}.

First-principles electronic structure approaches have been very successful
for bulk ferroelectrics \cite{14,15,16,17,18,19,20,21} but the application
of first-principles band structure methods to surface properties is daunting
due to the giant computational burden. Here the first set of accurate
electronic structure calculations for periodic BaTiO$_3$ slabs are
presented. These results can be considered benchmark results for faster,
more approximate methods, and also give some insights into the electronic
structure of ferroelectric BaTiO$_3$ surfaces.

\section{Method}

We studied six to seven atomic layer (001) and (111) periodic slabs of BaTiO$%
_{3\text{ }}$containing 15-18 atoms in the periodic unit, with seven to six
layers of vacuum, using the Linearized Augmented Plane Wave method with
extra local orbitals (LAPW+LO). \cite{22} Both slabs were studied in
the ideal configuration, and the more stable (001) slab was studied
extensively with tetragonal ferroelectric distortions and with surface
relaxations. In each case the in-plane lattice constant was 7.57 bohr
(=4.006 \AA ). An ABO$_3$ perovskite slab can have two types of
terminations, AO (Type I) and BO$_2$ (Type II). Three configurations for the
(001) slab were studied, asymmetrically terminated (BaO and TiO$_2$
surfaces), and symmetrically terminated with BaO and TiO$_2$ surfaces,
respectively. The slabs in each case were repeated with periodic boundary
conditions. The (001) slab had a repeat length of six times the cubic
lattice constant, or 45.42 bohr (=24.035 \AA ); whereas for the (111) slab $%
a $=10.70 bohr $c$=26.2 bohr. The ideal slabs are illustrated in
figure\thinspace \ref{fig_structures}. The (001) slab has P4mm tetragonal
symmetry with eight space group operations, and the (111) slab is
rhombohedral with R3m symmetry. In order to obtain accurate energy
differences and surface energies, bulk calculations were also performed in
the same symmetry with a tripled supercell with the same k-points and
convergence parameters.

The ferroelectric distortion in the (001) slab used the experimental
tetragonal average displacements. The Ba's were displaced by 0.06 \AA , the
Ti's by -0.1122 \AA\ along {\it c }relative to the O(II)'s, and the O(I)'s
by 0.0288 \AA\ . For the asymmetrically terminated slab, displacements were
performed in both directions, with Ti displaced towards the Ba-O surface (+)
and towards the Ti-O surface (-). The tetragonal phase in bulk BaTiO$_3$
does not consist primarily of displacements towards the perovskite cube
faces, as we are studying here, but rather displacements towards the cube
diagonals, and the tetragonal structure is a dynamical average with hopping
among four (111) directions. Such disordered configurations are well beyond
the possibilities of present computations for slabs, and must await future
faster and probably more approximate methods. Nevertheless, the general
effects of a surface on ferroelectric properties can be illuminated by the
present study.

The surface layers were relaxed using the LAPW forces \cite{23} and a
quasi-Newton method for the ideal (001) Ba-Ti terminated slab, the
ferroelectric (+) Ba-Ti terminated slab, and the ferroelectric Ba-Ba
terminated slab.

The convergence parameter RK$_{\max }$ was set to 7.0; Table II shows the
muffin tin radii. A 4$\times $4$\times $2 special k-point mesh was used
which gives 3 k-points for the (001) slab and 10 k-points for the (111)
slab. The ferroelectric Ba-Ti slab was also converged with a 6$\times $6$%
\times $2 mesh (6 k-points) and the total energy for the 15 atom unit cell
only differed by 0.2 mRyd. The matrix order (number of basis functions) for
the slabs was about 2900 for the first set of muffinb tin radii (see Table II) and 3800
for the second set. These calculations are extremely computationally
intensive. Not only was each iteration in the self-consistent cycle time
consuming but an unusually large number of iterations were required to reach
self-consistency due to charge fluctuations across the slab set up by the
electric fields generated across the slab.

\section{Results and Discussion}

Table I shows the unrelaxed and relaxed (001) slab structures, and total
energies for all studied slabs are shown in Table II. Since the (111) slab
is found to be much more unstable than the (001) slab, most of the
computations and the discussion below concentrate on the (001) slab. Whereas
there is a ferroelectric instability, or double well, found for bulk BaTiO$%
_3 $, there is no ferroelectric instability for the slab. Rather the energy
increases with ferroelectric distortion due to the depolarization field that
forms at the surface. This is the same result that was obtained using an
ionic model for finite clusters. \cite{2} Furthermore, for the
asymmetrically terminated Ba-Ti slab, the energy is not symmetric with
respect to ``+'' and ``-'' ferroelectric distortions, with a much lower
energy being obtained when the Ti atoms are moved so that the partially
coordinated surface Ti atoms move in towards the slab (``+'' distortion in
Table II).

\subsection{Relaxation}

The surface relaxations can be described as a combination of relaxation of
the surface layer in towards the slab, and a dimpling of the layers. Here we
define the former by the motion of the midpoint between the z coordinate for
the surface cation and anion relative to that of the ideal interlayer
spacing, and the latter by half of the difference in z-coordinate of the
surface ions. In agreement with work on other oxides, the sign of the
rumpling is such that the surface cations move in towards the slab and the
anions move out (Table III). The sign of the effect is independent of the
slab termination or whether the slab is ferroelectric or ideal. The
magnitude of the rumpling is quite similar to that obtained experimentally
for SrTiO$_3$. \cite{10} However, the experiment did not resolve net
relaxation of the layer spacing, whereas we find the that the surface layer
spacing is contracted relative to that of the bulk spacing. This may be
partly due to the fact that we did not relax all of the atomic positions in
the bulk of the slab, but rather fixed the interlayer spacing at the
experimentally observed zero pressure value, and LDA gives a smaller lattice
constant than experiment for the bulk. The relaxations are energetically
significant as is shown in Table II: 0.032 Ryd for the ideal (001) Ba-Ti
terminated slab, 0.016 Ryd for the ferroelectric Ba-Ti terminated slab, and
0.026 Ryd for the ferroelectric Ba-Ba terminated slab. For the Ba-Ti slab
the relaxation energy per surface is about 16\% of the cleavage energy.

\subsection{Surface energies}

Only the symmetrically terminated ideal slabs are expected to give accurate
estimates of the surface energy, because the asymmetrically terminated and
ferroelectric slabs have potential gradients (net polarization or
macroscopic electric field $E_{\text{mac}}$) which adds a term to the energy
proportional to $E_{\text{mac}}^2$. It is necessary to consider both the BaO
and TiO$_2$ terminated slabs together, since neither alone has bulk
stoichiometry, and thus its energy cannot be compared with bulk. Thus if we
add the energies of the BaO terminated slab and the TiO$_2$ terminated slab,
we get -128935.8994 Ryd for 7 BaTiO$_3$ units and 4 surfaces (2 BaO and 2 TiO%
$_2$). We subtract 7 times the bulk BaTiO$_3$ energy per cell (-128936.1703
Ryd) and get 0.2709 Ryd for 4 surfaces, or an average of 0.0677 Ryd/surface,
which gives 0.0574 eV/\AA $^2$ (=2712 erg/cm$^2$ = 0.92 J/m$^2$). For
comparison, the unrelaxed surface energy for Cr$_2$O$_{3\text{ }}$is
estimated to be 2000 erg/cm$^2$ for the relaxed surface \cite{24}. The
surface energy of SiO$_2$ glass is much lower, about 300 erg/cm$^2$. We have
not relaxed the symmetric TiO$_2$ terminated slab, but if its relaxation
energy is similar to that of the three slabs we did relax, relaxation would
lower the surface energy by about 12\%. The very high surface energy for
BaTiO$_{3\text{ }}$explains why it does not cleave easily, but rather
fractures. Real BaTiO$_{3\text{ }}$surfaces are probably highly defective
and measured surface energies may be therefore lower.

\subsection{Electronic structure}

Fig.\thinspace \ref{fig_band1} shows the calculated band structures the band
structure of the unrelaxed asymmetrically terminated Ba-Ti slab, the slab
with surface layers relaxed, and for bulk BaTiO$_3$ folded into the tripled
Brillouin zone of the slab. The relaxation makes small changes in the band
structure, but they are difficult to see on the scale shown. The top valence
band and bottom conduction band of the slab are interface states. The nature
of these states is clearer from Fig.\thinspace \ref{fig_band2} where the
band structures for the unrelaxed symmetrically terminated BaO and TiO$_2$
slabs are shown. The occupied interface state is due to the TiO$_2$
surfaces, and the interface state for the conduction band is primarily due
to the BaO surface.

The density of states was computed using the tetrahedron method with Fourier
interpolation \cite{25} from a k-point mesh of 8$\times $8$\times $2 to a 16$%
\times $16$\times $4 mesh in the full Brillouin zone. Fig.\thinspace \ref
{fig_BaO_dos} shows the density of states for the ferroelectric BaO
symmetrically terminated slab. There are Ba p -O p hybrid occupied surface
states, and an unoccupied O p band also comes down about 0.5 eV lower than
the bulk band. The Ba p states look quite different on the surface than in
the bulk, and in fact appear to be significantly less hybridized than the
bulk Ba, indicated than Ba is even more of a perfect 2+ ion on the surface
than in the bulk. The main valence band states on the surface O are
significantly less bound than the bulk O, with more weight at the top of the
valence band. Fig.\thinspace \ref{fig_titi_dos} shows that the occupied
surface states on TiO$_2$ surfaces have primarily O p character, with some
hybridization, interestingly, with the neighboring Ba in the next layer.
Fig.\thinspace \ref{fig_ferrodos} shows the density of states for the
unrelaxed and relaxed ferroelectric BaO symmetrically terminated slab. The
ferroelectric displacement sets up a potential gradient across the slab (a
net macroscopic field) that shifts the bands as a function of position in
the slab. In fact, the experimental ferroelectric displacement generates a
potential gradient large enough to make the slab metallic even for the thin
slabs studied here. The Ti d-states at the bottom of the slab become
partially occupied and the O p and Ba p states become partially empty at the
other end of the slab. Most interestingly, the bands do not move rigidly. It
appears that the O 2p states really do not want to be partially occupied,
and these states ``pile up'' against the Fermi level. The fact that this is
energetically preferred rather than allowing the oxygen to become partially
charged and highly nonspherical indicates how strongly the O atoms are
stabilized in the closed shell configuration, even on the surface of a slab.
What is so surprising about this is that O$^{2-}$ is not even stable in the
free state; it is only stabilized due to the Madelung (electrostatic) field
from the rest of the crystal. This is not an theoretical artifact; free O$%
^{2-}$ is unstable and decays to O$^{-}$ $+$ e$^{-}$. It appears that this
stabilization is strong enough even on the BaO perovskite surface. The
charge transfer across the slab screens partially screens the field, so that
the gap is just closed, as is discussed further below.

In band theory as applied here, there is a constant Fermi level and states
are occupied up to it. In an ideal insulating slab or crystal in a field,
however, the situation would be different and the Fermi level would vary
with macroscopic position and charge would not flow since states would not
communicate over macroscopic distances except by tunneling which is very
slow. This is similar to a hydrogen atom in an electric field; the ground
state for an infinitesimal field is for the electron to be stripped off, but
this does not happen even over long times due to a long lifetime for the
metastable state. In a real slab or crystal, however, there are extrinsic or
thermally induced defects so the conductivity is finite. Charge flow thus
would occur and the results would be similar to what we obtain here in the
static limit.

\subsection{Analysis of potential and charge densities}

Figure \ref{fig_den}a shows the charge density of the asymmetrically
terminated ideal slab with overlapping spherical ions subtracted. On the TiO$%
_2$ surface the O p charge density, instead of dangling, bonds back to the
surface Ti's. In other words, the surface charge density is self-healing.
The bonds between the O and Ti are clearly evident. Most interesting is the
collapse of the surface Ti bond. Instead of dangling the charge moves back
onto the Ti and the Ti-O surface bonds. This self-healing leaves the surface
highly reactive but stable (i.e. it doesn't reconstruct), and is probably
responsible for the utility of BaTiO$_3$ as a substrate for epitaxial growth
and for surface catalysis. The Ba-O surface shows much less difference from
spherical ions, and is found to be highly ionic. These results suggest that
the Ti-O surface of BaTiO$_3$ is highly reactive due to the possibility of
covalent bonding on the surface, whereas any reactivity Ba-O surface is due
entirely to ionic bonding. The non-bonding O p surface state may also
enhance surface reactivity.

Figure \ref{fig_den}b shows the difference between the LAPW (001) slab
density and the self-consistent LAPW charge density for bulk periodic BaTiO$%
_3$. Large changes are seen on the surface Ti, and smaller changes on the
surface Ba and O. Most interestingly, the differences the interior Ti and O,
which are only one unit cell away from the surface, are almost identical to
bulk Ti and O. This shows that electronic perturbations due the surface are
screened very rapidly in the interior of a crystal. The screening is
accomplished primarily due to polarization and charge redistributions around
the surface Ti, and to a lesser extent, O ions.

In order to understand the polarization and depolarization fields of the
slab, we now examine the total potentials averaged in the x-y plane as a
function of z-coordinate in the slabs. Figure \ref{fig_BaO_potential} shows
the potential for the symmetric BaO terminated ideal and ferroelectric
slabs. There is no net field in the vacuum region in the ideal case, and the
field is a constant 0.015 Ryd/bohr in the vacuum in the ferroelectric case.
The results for the asymmetrically terminated slabs are much more
complicated, since a field is present in the vacuum even in the ideal case
(Fig.\thinspace \ref{fig_asym_field}). The existence of the field in this
case can be thought of as due to different work functions for the BaO and TiO%
$_2$ surfaces, but in any case it indicates that the slab has a net dipole
moment in spite of the fact that if one considers the nominal charges (2+,
4+ and 2- for Ba, Ti and O respectively), the TiO$_2$ and BaO planes would
be charge balanced and the ideal slab would have no net polarization. One
possible explanation of the result is that a surface charge develops due to
the surface states. This possibility was tested by comparing the charge
densities of the asymmetrically terminated slab with the symmetrically
terminated BaO and TiO$_{2\text{ }}$slabs, but no evidence for a surface
charge could be found. The implication is that even in the bulk the BaO and
TiO$_2$ planes do not carry the same charge.

The above analysis allows the field to be easily extracted from the vacuum
region, but does not easily allow extraction of the macroscopic field in the
slab since the potential is dominated by local contributions from the atoms.
We can extract macroscopic field effects in the slabs by examining deep core
states. In figure \ref{fig_BaO_core} the O 1s core levels are shown as a
function of position in the symmetrically terminated BaO slab. As expected
there is no net field in the ideal slab, but there is a nearly constant
macroscopic field as determined from the O 1s levels in the ferroelectric
slab. The picture in the asymmetrically terminated slab is again complicated
(Fig.\thinspace \ref{fig_asym_core}). The net field as seen in the center of
the asymmetrically terminated ideal slab is very small, in spite of the
large field in the vacuum. The field for the ferroelectric relative to the
ideal asymmetric slab is nearly constant, and is about 0.01 Ryd/bohr.

We can understand better the observed macroscopic fields by considering the
ideal case of a periodic slab of thickness $L_1$ separated by vacuum regions
of thickness $L_2$ with constant fields $E_1$ and $E_2$ in the two regions.
The electric displacements are given by $D_1=E_1+4\pi P_1$ in the slab and $%
D_2=E_2$ in the vacuum, where $P_1$ is the total polarization in the slab,
including the spontaneous polarization $P_s$ and the induced polarization $%
P_i=(\varepsilon _1-1)E_1/4\pi $. Now $D_1=D_2$ and $E_1L_1+E_2L_2=0$ since
the potential must be continuous. For the BaO terminated slab we get $%
E_1=0.005$ H/bohr and $E_2=-0.0073$ H/bohr which gives $L_1/L_2=1.45$, which
compares well with the simple consideration of 7 slab layers and 5 vacuum
layers, or $L_1/L_2=1.4$, indicating that our method of estimating $E_1$ and 
$E_2$ from the vacuum potential and the O 1s core levels is consistent.
Using the experimental dielectric constant $\varepsilon =5.24$ we get $P_s=$%
15 $\mu $C/cm$^2$, which is significantly lower than the experimental value
of 26 $\mu $C/cm$^2$. The screening caused by the metallic band structure is
most likely the major cause of this difference. This has important
implications, and shows that with finite conductivity, a ferroelectric will
self-screen the depolarization field, because of the effective closing of
the gap from one part of a ferroelectric domain to the other. Many of the
implications remain to be worked out.

Finally, we consider the magnitude of the depolarization energy, $W_D=\frac 1%
2\int D\cdot E$ d$V$ . For the ferroelectric symmetrically terminated BaO
slab we find $W_D=0.055$ Ryd ($=$ 0.75 eV), a large energy which easily
overcomes the energy lowering for the ferroelectric distortion in bulk BaTiO$%
_3$ of 0.015 eV per unit cell, or 0.053 eV for the equivalent number of
atoms as the slab contains in a periodic unit by a factor of about 14. Note
that both numbers scale up with size, so even macroscopic crystals must
break up into domains, or the depolarization field must be pacified by
formation of a surface space charge.

\section{Conclusions}

We have studied ferroelectric and ideal slabs of BaTiO$_3$ using the
full-potential all-electron LAPW method. Surface relaxations are
significant, and similar to what is observed for SrTiO$_3$. There are
significant surface states, especially on the TiO$_2$ surfaces. The dominant
surface state has O p character, and seems to be related to the collapse of
the dangling bonds back onto the surface Ti ions. The charge density in the
center of the slab is very close to the charge density in bulk BaTiO$_3$. We
find the asymmetrically terminated slabs have significant potential
gradients (electric fields) in the vacuum due to the different net charges
on BaO and TiO$_2$ planes. Most interestingly, we find that the
ferroelectric slabs are metallic due to potential shifts that are greater
than the band gap. The bands, however, do not shift in a rigid way; rather
the O 2p states pile up at the Fermi level showing that the O atoms prefer
to remain closed-shelled, even on the surface. The metallic nature of the
ferroelectric slabs screens the macroscopic polarization, so we find a
polarization smaller than the experimental bulk value. The depolarization
field is large and inhibits ferroelectricity. In order to recover a
ferroelectric state, this field must be screened or the crystal must break
up into domains.

\acknowledgments{Thanks to V. Heine, I. Mazin, R. Resta, and D. Vanderbilt
for helpful discussions and suggestions on how to interpret the slab
results. D. Vanderbilt suggested the study of symmetrically terminated slabs
which was very useful. This research is supported by the Office of Naval
Research. Computations were performed on the Cray J90 at the Geophysical
Laboratory.. }

\begin{table}[tbp]
\caption{Structures of (001) BaTiO$_3$ slabs (z coordinates shown only, in
lattice coordinates, $c = $24.035 \AA). Only relaxed coordinates are shown
for partially relaxed structures.}
\begin{tabular}{lllllllllll}
&  & ideal & relaxed & ferro - & ferro + & relaxed & ideal & ferro & relaxed
& ideal \\ 
atom x,y & termination & Ba-Ti & Ba-Ti & Ba-Ti & Ba-Ti & Ba-Ti & Ba-Ba & 
Ba-Ba & Ba-Ba & Ti-Ti \\ 
Ba 0,0 &  & 0 & 0.00715 & 0.0025 & -0.0025 & 0.00596 & 0 & -0.0025 & 0.00762
& 0 \\ 
Ba 0,0 &  & 0.16667 &  & 0.16917 & 0.16417 &  & 0.16667 & 0.16417 &  & 
0.16667 \\ 
Ba 0,0 &  & 0.33333 &  & 0.33583 & 0.33083 &  & 0.33333 & 0.33083 &  & 
0.33333 \\ 
Ba 0,0 &  &  &  &  &  &  & 0.5 & 0.50128 & 0.49110 &  \\ 
Ti 0.5,0.5 &  &  &  &  &  &  &  &  &  & -0.08333 \\ 
Ti 0.5,0.5 &  & 0.08333 &  & 0.088 & 0.07867 &  & 0.08333 & 0.07867 &  & 
0.08333 \\ 
Ti 0.5,0.5 &  & 0.25 &  & 0.25467 & 0.24533 &  & 0.25 & 0.24533 &  & 0.25 \\ 
Ti 0.5,0.5 &  & 0.41667 & 0.40868 & 0.42133 & 0.412 & 0.40910 & 0.41667 & 
0.412 &  & 0.41667 \\ 
O 0,0 &  &  &  &  &  &  & 0 & 0.00128 &  & -0.08333 \\ 
O 0.5,0.5 &  & 0 & 0.00544 & -0.00128 & 0.00128 & 0.00241 & 0.08333 & 0.08333
& 0.00276 & 0 \\ 
O 0,0 &  & 0.08333 &  & 0.08333 & 0.08333 &  & 0.1667 & 0.16795 &  & 0.08333
\\ 
O 0.5,0.5 &  & 0.16667 &  & 0.1653 & 0.16795 &  & 0.25 & 0.25 &  & 0.16667
\\ 
O 0,0 &  & 0.25 &  & 0.25 & 0.25 &  & 0.33333 & 0.33462 &  & 0.25 \\ 
O 0.5,0.5 &  & 0.33333 &  & 0.332 & 0.33462 &  & 0.41667 & 0.41667 &  & 
0.33333 \\ 
O 0,0 &  & 0.41667 & 0.41219 & 0.41667 & 0.41667 & 0.41312 & 0.5 & 0.50128 & 
0.49199 & 0.41667
\end{tabular}
\end{table}
\begin{table}[tbp]
\caption{Summary of total energies for BaTiO$_3$ slabs and bulk.}
\begin{tabular}{lllll}
& Termination & E Ryd/cell &  &  \\ 
symmetric bulk (3 (001) layers) \tablenotemark[1] &  & $-55258.3586$ &  & 
\\ 
symmetric bulk (2 (111) layers)\tablenotemark[1] &  & $-55258.3578$ &  &  \\ 
ideal (001) slab \tablenotemark[1] & Ba-Ti & $-55258.1189$ &  &  \\ 
ideal (111) slab\tablenotemark[1] & Ba-Ti & $-55257.4260$ &  &  \\ 
ferroelectric slab ($-$) \tablenotemark[1] & Ba-Ti & $-55258.0482$ &  &  \\ 
ferroelectric slab ($+$) \tablenotemark[1] & Ba-Ti & $-55258.1119$ &  &  \\ 
symmetric bulk (3 (001) layers)\tablenotemark[2] &  & $-55258.4348$ &  &  \\ 
ideal (001) slab \tablenotemark[2] & Ba-Ti & $-55258.2119$ &  &  \\ 
(001) slab relaxed \tablenotemark[2] & Ba-Ti & $-55258.2434$ &  &  \\ 
ferroelectric slab ($+$) \tablenotemark[2] & Ba-Ti & $-55258.2063$ &  &  \\ 
ferroelectric slab ($+$) relaxed \tablenotemark[2] & Ba-Ti & $-55258.2222$ & 
&  \\ 
ideal (001) slab \tablenotemark[2] & Ba-Ba & $-71674.0910$ &  &  \\ 
ferroelectric (001) slab \tablenotemark[2] & Ba-Ba & $-71674.0495$ &  &  \\ 
ferroelectric (001) slab relaxed \tablenotemark[2] & Ba-Ba & $-71674.0758$ & 
&  \\ 
ideal (001) slab \tablenotemark[2] & Ti-Ti & $-57261.8084$ &  & 
\end{tabular}
\tablenotetext[1]{Muffin tin radii: Ba=2.3 bohr, Ti=1.75 bohr,O=1.75} %
\tablenotetext[2]{Muffin tin radii: Ba=2.3 bohr, Ti=1.59938 bohr,O=1.59938}
\end{table}

\begin{table}[tbp]
\caption{Summary of relaxations of surface layers in \AA.}
\begin{tabular}{llllll}
&  & Ba-Ti ideal & Ba-Ti ferro + & Ba-Ba ferro & SrTiO$_3$\tablenotemark[2]
\\ 
& relaxation of Ba-O layer & 0.15 & 0.12 & 0.14 & -0.08 \\ 
& change in dimpling of Ba-O layer & 0.02 & 0.088 \tablenotemark[1] & 0.10 %
\tablenotemark[1] & 0.08 \\ 
& relaxation of Ti-O layer & 0.15 & 0.078 & 0.23 & -0.05 \\ 
& change in dimpling of Ti-O layer & 0.08 & 0.0078 & 0.011 & 0.05
\end{tabular}
\tablenotetext[1]{Sign of dimpling changes relative to ferroelectric
distortion.} \tablenotetext[2]{Ref.~\cite{10}}
\end{table}

\begin{figure}[tbp]
\caption{Structures of slabs studied. (a) (111) slab (b) (001) slab. Blue is
oxygen, green is Ba, and red is Ti. Ti-O bonds are shown in black.}
\label{fig_structures}
\end{figure}

\begin{figure}[tbp]
\caption{Band structures for (a) undistorted bulk BaTiO$_3$ folded into the
Brillouin zone for the three layer slab, and (b) for the assymetrically
terminated slab with BaO and TiO$_{2\text{ }}$surfaces. The bands for the
unrelaxed ideal slab are shown as solid lines, and for the slab with relaxed
surfaces are dashed.}
\label{fig_band1}
\end{figure}

\begin{figure}[tbp]
\caption{Band structures for symmetric (a) BaO and (b) TiO$_2$ terminated
ideal slabs. The highest valence band in (b) is a surface state with O 2p
character, and the lowest conduction band in (a) is also a surface state
with primarily O 2p character.}
\label{fig_band2}
\end{figure}

\begin{figure}[tbp]
\caption{Partial density of states for the symmetric ideal BaO terminated
slab. Note the different scales for Ba p and Ti d. The Fermi level is shown
as the vertical dashed line. The amount of Ti hybridization with O 2p is
about constant through the slab, but the Ba p hybridization is lower on the
surface than in the bulk. The O 2p bands are correspondingly wider in the
bulk than on the surface. The surface is more ionic than the bulk.}
\label{fig_BaO_dos}
\end{figure}

\begin{figure}[tbp]
\caption{Partial density of states for the symmetric ideal TiO$_2$
terminated slab. There is a significant surface state with O 2p character as
was shown in fig.\thinspace \ref{fig_band2}. The surface seems slightly more
ionic than the interior of the slab, and the Ti d states are slightly less
bound.}
\label{fig_titi_dos}
\end{figure}

\begin{figure}[tbp]
\caption{Partial density of states for the ferroelectric symmetrically BaO
terminated slab. The Ti displacements were towards the bottom of the slab
(corresponding to the bottom panels). The solid line is for the ideal
ferrolectric slab with the experimental tetragonal displacement pattern, and
the dashed line is for the slab with relaxed surface atoms. The slab is
metallic, with conduction bands crossing the Fermi level at the bottom of
the slab and O 2p states and Ba p states crossing the Fermi level at the top
of the slab. Note that the bands are displaced due to the ramp potential
across the slab, but the band shifts are not rigid. Rather the O 2p density
piles up at the Fermi level at the top of the slab, indicating that the O
ion wants to remain closed shell. Only a small amount of charge transfer
occurs in order to screen the macroscopic field to the point that the gap is
just closed.}
\label{fig_ferrodos}
\end{figure}

\begin{figure}[tbp]
\caption{Charge density for BaTiO$_3$ slab. (a) Deformation charge density
for BaTiO$_3$. The image shows the difference in charge density between the
self-consistent LAPW charge density and overlapping PIB ions. Green
represents no change in density (i.e. identical to the ionic crystal).
Partially charged ions, Ba$^{1.63+}$, Ti$^{3.26+}$, O$^{1.63-}$ were used to
generate the ionic charge density. Note the ``healing'' of the surface Ti
dangling bond. It collapses back onto the Ti. (b) Difference in charge
density between the ideal slab and bulk BaTiO$_3$. Even though the slab is
only three layers thick, the central layer has almost an identical charge
density to bulk BaTiO$_3$, although there are still some small differences
on the Ba.}
\label{fig_den}
\end{figure}

\begin{figure}[tbp]
\caption{Total potential averaged in the x-y plane as a function of z for
the symmetric ideal and ferroelectric BaO terminated slabs. The average
field is zero is zero in the vacuum between slabs in the ideal case, and is
0.39 eV/\AA\ in the ferroelectric case. The experimental tetragonal
displacement pattern is used.}
\label{fig_BaO_potential}
\end{figure}

\begin{figure}[tbp]
\caption{Electric field shown only in the vacuum region for the
asymmetrically terminated slabs. Abscissa is for the whole periodic slab
and vacuum. The field is not shown in the slab interior since it
oscillates widely on this scale.}
\label{fig_asym_field}
\end{figure}

\begin{figure}[tbp]
\caption{Core levels for symmetric BaO terminated slabs. (a) O 1s core
energies as a function of layer (z coordinate) in slabs for ideal and
ferroelectric slab. (b) Macroscopic electric field derived from changes in
core levels.}
\label{fig_BaO_core}
\end{figure}

\begin{figure}[tbp]
\caption{Core levels for asymmetrically terminated slabs. (a) O 1s core
energies as a function of layer (z coordinate) in slabs for ideal and
ferroelectric slabs, relaxed and unrelaxed. (b) Macroscopic electric field
derived from changes in core levels.}
\label{fig_asym_core}
\end{figure}

\end{document}